\documentclass[10pt, twocolumn, superscriptaddress,preprintnumbers,showpacs,showkeys]{revtex4}
\usepackage{epsf} 
\usepackage{amsmath}
\usepackage{amssymb}
\usepackage{epsfig}
\usepackage{latexsym}
\usepackage{amsfonts}
\usepackage{graphicx}%
\usepackage{varioref}
\usepackage{ifthen}
\begin{document}
\parindent 0mm 
\setlength{\parskip}{\baselineskip} 
\thispagestyle{empty}
\pagenumbering{arabic} 
\setcounter{page}{1}
\mbox{ }

\preprint{UCT-TP-273/08}

\title{Scalar radius of the pion in the Kroll-Lee-Zumino renormalizable theory}
\vspace{.1cm}
\author{C. A. Dominguez}
\affiliation{Centre for Theoretical Physics and Astrophysics, University of
Cape Town, Rondebosch 7700}
\affiliation{Department of Physics, Stellenbosch University, Stellenbosch 7600, South Africa}
  
\author{M. Loewe}
\affiliation{Facultad de F\'{i}sica, Pontificia Universidad Cat\'{o}lica de Chile, Casilla 306, Santiago 22, Chile}

\author{B. Willers}
\affiliation{Centre for Theoretical Physics and Astrophysics, University of
Cape Town, Rondebosch 7700}
\date{\today}

\begin{abstract}
The Kroll-Lee-Zumino renormalizable Abelian quantum field theory of pions and a massive rho-meson is used to calculate the scalar radius of the pion at next to leading (one loop) order  in perturbation theory. Due to renormalizability, this determination involves no free parameters. The result is $\langle r^2_\pi\rangle_s = 0.40\;  \mbox{fm}^2$. This value gives for $\bar{\ell}_4$, the low energy constant of chiral perturbation theory, $\bar{\ell}_4 = 3.4$, and $F_\pi/F = 1.05$, where F is the pion decay constant in the chiral limit. Given the level of accuracy in the masses and the $\rho\pi\pi$ coupling, the only sizable uncertainty in this result is due to the (uncalculated)  NNLO contribution.
\end{abstract}

\maketitle
\noindent
The pion matrix element of the QCD scalar operator $J_S = m_u \bar{u}u + m_d \bar{d}d$ defines the scalar form factor of the pion \cite{FS}
\begin{equation}
\Gamma_\pi(q^2) = \langle \pi(p_2) | J_S | \pi(p_1) \rangle \;,
\end{equation}
where $q^2 = (p_2 - p_1)^2$. The associated quadratic scalar radius
\begin{equation}
\Gamma_\pi(q^2) = \Gamma_\pi(0) \Bigl[ 1 + \frac{1}{6} \langle r^2_\pi\rangle_s q^2 +...\Bigr],
\end{equation}
plays a very important role in chiral perturbation theory \cite{REVCPT}, as it fixes $\bar{ l }_4$, one of the low energy constants of the theory, through the relation
\begin{equation}
\langle r^2_\pi\rangle_s = \frac{3}{8 \pi^2 F_\pi^2}\;\Bigl[ \bar{\ell}_4 - \frac{13}{12} +
O (M_\pi^2) \Bigr]\;,
\end{equation}
where $F_\pi = 92.4 \;\mbox{MeV}$. The low energy constant $\bar{\ell}_4$, in turn, determines the leading contribution in the chiral expansion of the pion decay constant, i.e.
\begin{equation}
\frac{F_\pi}{F} = 1 + \left(\frac{M_\pi}{4 \pi F_\pi}\right)^2 \;\bar{\ell}_4 + O(M_\pi^4),
\end{equation}
where F is the pion decay constant in the chiral limit.
For this reason considerable effort has been devoted over the years to the determination of $\langle r^2_\pi\rangle_s$ from $\pi \pi$ scattering data together with a variety of theoretical tools (for some recent work see \cite{V1}-\cite{OLLER}). Current values \cite{OLLER} appear to  converge inside the range $\langle r^2_\pi\rangle_s \simeq 0.5 - 0.7 \;\mbox{fm}^2$
which translates into $\bar{\ell}_4 \simeq 4.0 - 5.1$, and $F_\pi/F \simeq 1.06 - 1.07$. Lattice QCD results \cite{LQCD} span the wide range  $\bar{\ell}_4 = 3.6 - 5.0  $, although results with the smaller errors cluster around $\bar{\ell}_4 \simeq 3.8 - 4.5$.\\

In this paper we present a next to leading order calculation of $\langle r^2_\pi\rangle_s$ in the framework of the Kroll-Lee-Zumino (KLZ) renormalizable Abelian gauge theory of charged pions and  a massive neutral vector meson \cite{KLZ}. This theory provides the quantum field theory justification for the Vector Meson Dominance (VMD) ansatz \cite{VMD}. It also provides  a quantum field theory platform to compute corrections to VMD systematically in perturbation theory. A determination in this framework of the electromagnetic form factor of the pion in the time-like \cite{GK} as well as the spacelike region \cite{CAD1}, at the one-loop level, which is in excellent agreement with data supports this assertion. In fact, due to the relative mildness of the  $\rho\pi\pi$ coupling constant, and the  presence of loop suppression factors, the perturbative expansion appears well behaved in spite of the strong coupling nature of the theory. The KLZ Lagrangian is given by

\begin{eqnarray}
\mathcal{L}_{KLZ} &=& \partial_\mu \phi \, \partial^\mu \phi^* -  M_\pi^2 \,\phi \,\phi^* - \tfrac{1}{4}\, \rho_{\mu\nu} \,\rho^{\mu\nu} 
+ \tfrac{1}{2}\, M_\rho^2\, \rho_\mu \,\rho^\mu \nonumber\\ [.3cm]
&+&g_{\rho\pi\pi} \rho_\mu J^\mu_\pi\ + g_{\rho\pi\pi}^2 \;\rho_\mu\; \rho^\mu \;\phi \;\phi^*   \;,
\end{eqnarray}
where $\rho_\mu$ is a vector field describing the $\rho^0$ meson ($\partial_\mu \rho^\mu = 0$), $\phi$ is a complex pseudo-scalar field describing the $\pi^\pm$ mesons, $\rho_{\mu\nu}$ is the usual field strength tensor: $\rho_{\mu\nu}  = \partial_\mu \rho_\nu - \partial_\nu \rho_\mu$, and $J^\mu_\pi$ is the $\pi^\pm$ current: $J^\mu_\pi  = i \phi^{*} \overleftrightarrow{\partial_\mu} \phi$.
It should be stressed that in spite of the explicit presence of the $\rho^0$ mass term above, the theory is renormalizable because the neutral vector meson is coupled only to a conserved current \cite{KLZ}.\\

In Fig.1 and in Fig.2 we show, respectively, the leading order, and the next to leading order contributions to the scalar form factor Eq.(1). The cross indicates the coupling of the scalar operator to two pions. There is still another triangle graph with two rho-mesons coupled to the scalar current $J_S$. However, since the scalar form factor vanishes identically in the chiral limit, two rho-mesons would have to couple to $J_S$ through two pions (a coupling present in Eq.(5)). This transforms this term into a two-loop $\cal{O}$  ($g_{\rho\pi\pi}^4$) contribution, which is beyond the scope of the present work.
\begin{figure}
[ht]
\begin{center}\includegraphics[width=2.0 in]
{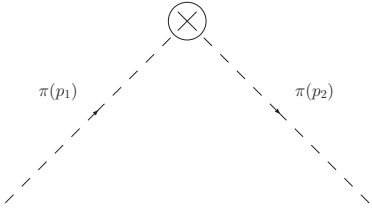}
\caption{Leading order contribution to the scalar form factor of the pion. The cross indicates the coupling of the scalar current to two pions.}
\end{center}
\end{figure}

\begin{figure}
[ht]
\begin{center}
\includegraphics[width=2.0 in]
{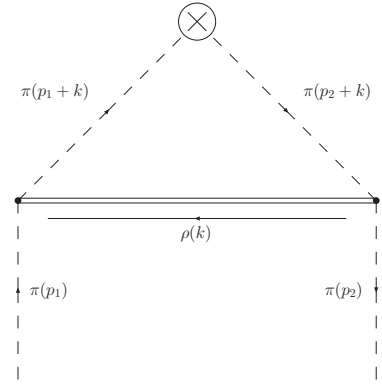}
\caption{Next to leading order contribution to the scalar form factor.}
\end{center}
\end{figure}

Using the Feynman propagator for the $\rho$-meson \cite{Hees}-\cite{Quigg}, and in $d$-dimensions, the unrenormalized vertex in Fig. 2 is given by
\begin{widetext}
\begin{equation}
\widetilde{G}(q^2)  =  g_{\rho\pi\pi}^2 (\mu^3)^{2 - \frac{d}{2}}  
\int \frac{d^dk}{(2 \pi)^d}
\frac{ (2 {p_1} + k) \cdot (2 {p_2} + k)} 
{[({p_1} + k)^2 - M_\pi^2 + i \varepsilon] [({p_2} + k)^2 - M_\pi^2 + i \varepsilon] (k^2 - M_\rho^2 + i \varepsilon)} \;,
\end{equation}
\end{widetext}

where we omitted the overall normalization $\Gamma_\pi(0)$.
Using standard procedures (for details of a similar calculation see \cite{CAD1}) the function $\widetilde{G}(q^2)$ in dimensional regularization is
\begin{eqnarray}
\widetilde{G} (q^2) &=&  -2\;\frac{g_{\rho\pi\pi}^2}{(4 \pi)^2} \left( \mu^2 \right)^{(2 - \frac{d}{2})} \int_0^1 dx_1\int_0^{1-x_1} dx_2 \nonumber \\ [.3cm]
&\times&
\left\{ \frac{2}{\varepsilon} - \ln \left( \frac{\Delta(q^2)}{\mu^2}\right) 
- \frac{1}{2} - \gamma + \ln (4 \pi)  \right. \nonumber \\ [.3cm]
&+& \left. \frac{1}{2 \Delta(q^2)} \left[M_\pi^2 (x_1 + x_2 - 2)^2 
\right. \right.\nonumber \\ [.3cm]
&-& \left. \left. q^2 (x_1 x_2 - x_1 - x_2 +2)\right] + O(\varepsilon) \phantom{\frac{1}{1}} \right\} \;,
\end{eqnarray}
where $\Delta(q^2)$ is defined as
\begin{equation}
\Delta(q^2) = M_\pi^2 (x_1+x_2)^2 + M_\rho^2(1-x_1-x_2) - x_1 x_2 q^2 \;.
\end{equation}
In the $\overline{MS}$ scheme, and renormalizing the vertex function at the point $q^2=0$ we obtain 
\begin{eqnarray}
&&G(q^2) - G(0) = - 2\;  \frac{g_{\rho\pi\pi}^2}{(4 \pi)^2} \int_0^1 dx_1 \int_0^{1-x_1} dx_2  \nonumber \\ [.3cm]
&\times& \left\{ \ln \left( \frac{\Delta(q^2)}{\Delta(0)}\right)  + \frac{1}{2}  \left[ M_\pi^2 (x_1 + x_2 - 2)^2 \left(\frac{1}{\Delta(q^2)}
\right. \right. \right. \nonumber \\ [.3cm]
&-&  \left. \left.  \left.  \frac{1}{\Delta(0)}\right)
- \frac{q^2}{\Delta(q^2)} (x_1 x_2 - x_1 - x_2 +2) \right] \right\} \;,
\end{eqnarray}
with the scalar form factor being given by
\begin{equation}
\Gamma_\pi(q^2) = \Gamma_\pi(0) \left[ 1 + G(q^2) - G(0)\right] \;.
\end{equation}
Details of the standard renormalization procedure for the fields, masses and coupling may be found in \cite{CAD1}. From Eq.(9) we compute the scalar radius with the result
\begin{eqnarray}
\langle r^2_\pi\rangle_s &=& \frac{12}{(4 \pi)^2}\; g^2_{\rho\pi\pi} \;\int_0^1 dx_1 \int_0^{1-x_1} dx_2 \;\frac{1}{\Delta(0)} \nonumber \\ [.3cm]
&\times& \left\{x_1 x_2 \left[ 1 - \frac{M_\pi^2}{2\Delta(0)}  (x_1+x_2-2)^2\right] \right. \nonumber \\ [.3cm]
&+& \left. \frac{1}{2}(x_1 x_2 - x_1 - x_2 +2) \right\} \;.
\end{eqnarray}

A numerical evaluation of this equation gives the result
\begin{equation}
\langle r^2_\pi\rangle_s = 0.4 \mbox{ fm}^2 \; ,
\end{equation}
where we used $g^2_{\rho\pi\pi} = 36.0 \pm 0.2$ from the measured width of the $\rho$ \cite{PDG}. The error in this coupling, as well as in the masses, has negligible impact on the radius at the level of precision given in Eq.(12). The main uncertainty in this determination stems from the uncalculated NNLO (two-loop) contribution. Using Eqs.(3) and (4) to leading order,  the result above translates into
\begin{equation}
\bar{\ell}_4 = 3.4 \;, 
\end{equation}
and 
\begin{equation}
F_\pi/F = 1.05 \;.
\end{equation}
The result for the radius in this framework is somewhat smaller than current values obtained from $\pi \pi$ scattering \cite{V1}-\cite{OLLER}, although it agrees with some of the lattice QCD results  \cite{LQCD}. It should be mentioned that in the framework of KLZ the electromagnetic square radius of the pion at NLO is \cite{CAD1} $\langle r^2_\pi\rangle_{EM} = 0.46\; \mbox{fm}^2$, to be compared with the experimental value \cite{RADIUSEM} $\langle r^2_\pi\rangle_{EM} = 0.439 \pm 0.008\;\mbox{fm}^2$. In the electromagnetic case NLO refers to the correction  to the tree-level result of single $\rho$-dominance $\langle r^2_\pi\rangle_{EM}|_{LO} = 0.39 \;\mbox{fm}^2$. Hence this correction is relatively large, and in the right direction. In the present application the equivalent of $\rho$-dominance is absent, as there is no elementary sigma field in the KLZ Lagrangian. One would have to resort to e.g. the linear sigma model as in \cite{GL}, but then there is no $\rho$ field in the model. An attempt to enlarge the KLZ theory to accommodate a sigma field does not seem  a useful proposition. In fact, scalar meson dominance is probably too simplistic to be able to account for the rich and complex structure of the $J^P = 0^+$ channel.
We find that the result obtained here for the scalar radius of the pion provides additional support for the KLZ theory as a viable platform to compute corrections to VMD systematically in perturbation theory.\\

{\bf Acknowledgements}\\

The authors wish to thank to Heiri Leutwyler for a valuable discussion, and for his comments on the manuscript. This work has been supported in part by FONDECYT 1051067, 7070178, and by Centro de Estudios Subatomicos (Chile), and by NRF (South Africa).


\begin{thebibliography}{99}
\bibitem{FS} T.N. Truong, and R.S. Willey, Phys. Rev. D 40,3635 (1989); J.F. Donoghue, J. Gasser, and H. Leutwyler, Nucl. Phys. B 343, 341 (1990).

\bibitem{REVCPT} For reviews see e.g. S. Scherer, Adv. Nucl. Phys. 27, 277 (2003); J. Gasser, Lect. Notes Phys. 629, 1 (2004).

\bibitem{V1} B. Moussallam, Eur. Phys. J. C 14, 111 (2000); G. Colangelo, J. Gasser,and  H. Leutwyler, Nucl. Phys. B 603, 125 (2001); B. Ananthanarayan, I. Caprini, G. Colangelo, J. Gasser, and H. Leutwyler, Phys. Lett. B 602, 218 (2004); F.J. Yndurain, Phys. Lett. B 612, 245 (2005).

\bibitem{OLLER} J.A. Oller, and L. Roca, Phys. Lett. B 651, 139 (2007).

\bibitem{LQCD} For a recent review of the various determinations see e.g. S. Necco, PoS LAT 2007:021,2007, and arXiv:0710.2444.

\bibitem{KLZ} N.M. Kroll, T.D. Lee, and B. Zumino, Phys. Rev. 175 (1967) 1376; J.H. Lowenstein, and B. Schroer,Phys. Rev. D 6 (1972) 1553.

\bibitem{VMD} J.J. Sakurai, Ann. Phys. (N.Y.) 11 (1960) 1; {\it ibid.} Currents and Mesons, University of Chicago Press (1969).

\bibitem{GK} C. Gale, and J. Kapusta, Nucl. Phys. B 357 (1991) 65.
\bibitem{CAD1} C.A. Dominguez, M. Loewe, J.I. Jottar, and B. Willers, Phys. Rev. D 76, 095002 (2007). This paper has a misprint in Eq.(15) (the sign of the first term in curly brackets should be negative), with the remaining equations being correct. The electromagnetic square radius of the pion quoted in the paper is incorrect; the correct value is $\langle r^2_\pi\rangle_{EM} = 0.46\, \mbox{fm}^2$, in much better agreement with data than naive (single $\rho$) VMD.

\bibitem{Hees} H. van Hees, hep-th/0305076 (unpublished); H. Ruegg, and M. Ruiz-Altaba, Int. J. Mod. Phys. A 19 (2004) 3265.

\bibitem{Quigg} C. Quigg, Gauge Theories of Strong, Weak, and Electromagnetic Interactions, Benjamin (1983).

\bibitem{PDG} Review of Particle Physics, Particle Data Group, J. Phys. G: Nucl. Part. Phys. 33, 1 (2006).

\bibitem{RADIUSEM} NA7 Collaboration, S. R. Amendolia et al., Nucl. Phys. B 277, 168 (1986).

\bibitem{GL} J. Gasser, and H. Leutwyler, Ann. Phys. (N.Y) 158, 142 (1984).

\end{thebibliography}
\end{document}